# Amplified Spontaneous emission from optical fibers containing anisotropic morphology CdSe/CdS quantum dots under CW excitation

**Palash Kusum Das[1], Nishant Dhiman[2], Siva Umapathy[1,2], Frédéric Gérôme [3], Asha Bhardwaj[1*]**

## Affiliations

**1. Instrumentation and Applied Physics, Indian Institute of Science, Bengaluru, India**

**2. Inorganic and Physical Chemistry, Indian Institute of Science, Bengaluru, India,**

**3. XLIM-UMR 7252, University of Limoges/CNRS, 123 ave Albert Thomas, 87060 Limoges CEDEX, France**

## Corresponding author

Correspondence to: Asha Bhardwaj

## Abstract

Fibers lasers is a field which is typically dominated by rare earth ions as gain material in the core of a silica optical waveguide. Due to their specific emission wavelengths, rare- earth doped fiber lasers are available only at few pre-defined wavelengths. However, Quantum Dots (QDs) are materials, which shows tunable emission with change in size and composition. Due to such tunability, QDs seem to be promising candidates for obtaining fiber lasers at a spectrum of wavelengths which are not possible using rare earth ions. To replace rare earth ions with QDs, it is of paramount importance that QDs show signatures of optical gain. Here, we report synthesis of asymmetric pod-shaped CdSe/CdS QDs, which demonstrate efficient gain through pumping. The intrinsic gain properties of the QDs have been evaluated through Transient Absorption Spectroscopy. Later, the exquisite QDs are used to fabricate specialty fibers from which ASE has been obtained by using CW laser pump at Room Temperatur. Finally, Stability of the emission signal has been by studying photobleaching and controlling the concentration of the QDs

## Introduction

For the past two decades, fibre lasers have dominated the commercial laser market due to their versatility, output power quality, and small size[38]. The trivalent rare-earth ion-doped silicate glass active zones in the fibre laser sources include Er3+, Tm3+, Nd3+, and Yb3+ which operate at predetermined wavelengths that correspond to their respective 4f-4f transitions[1-6,9-11]. Their definite output wavelengths, limit the employment of fiber lasers in a wide range of fields.

However, Quantum dots (QDs) on the other hand, due to the quantum confinement of charge carriers are highly emitting materials which manifest size and composition-dependent tunability of optical properties. Such tunability with size and composition due to quantum confinement effects and discrete band structure facilitates engineering their emission wavelengths and achieving amplification. Their application[13-19] arena increases due to their simpler, low-cost chemical processability and high quantum yields (QYs).

Further, CdSe nanocrystals with epitaxially grown CdS shell (CdSe/CdS QDs) have been proven to be excellent candidates for obtaining light amplification and lasing due to their small conduction band offsets[57] and dependence of emission characteristics on morphology and thickness of CdS shell[51]. It is also observed that adjusting wavefunction overlap and minimizing auger recombination by tailoring the morphology aids in achieving optical gain. Additionally, asymmetrically strained structures create large light-heavy hole splitting, lowering the band edge degeneracy and thus dramatically lowering the lasing threshold[52-54]. Synthesis of heterostructures with complex anisotropic shapes (such as dot-in-rods, rod-in-rods, tetrapods, and octapods) is still in its infancy with regard to the most recent QD lasing technology [55].

Keeping in mind, the tunable nature of CdSe/CdS QDs with size and gain enhancement for anisotropic morphology, these QDs would be an excellent choice for the replacement of traditional rare-earth ions for realizing fiber lasers at unconventional wavelengths. In recent years, optical properties of QD doped optical fibers have been extensively studied for their potential applications in fiber-based amplifiers, lasers, and sensors. Hreibi et al. reported the fabrication and characterization of a PbSe QDs doped liquid-core fibres[20]. Cheng et al. reported a PbSe QD-doped fibre laser (QDFL) with a ring resonator using colloidal PbSe QDs as gain media[21]. Lei Zhang developed a theoretical model of PbSe doped fibres[22,23] and also investigated the optical properties of PbSe/CdSe QDs doped fibres [24]. Bahrampour et al. investigated PbSe doped fibre in a single-mode condition[25].

Also, numerous works has been reported on QD ASE (Amplified Spontaneous Emission) and lasing in different cavity modes. Earlier, most of the ASE that has been obtained from colloidal QD medium are obtained in closely packed thin films[39-41] to study intrinsic gain property without the effect of Lasing cavity. Also, most of the studies that have been done are based on femto second (fs)[40,43] or ns pumping (quasi-CW)[44,50] using a Cryostat setup[45]. Although, there has been reports on investigation of QDs containing optical fibers, but not much work has been done in obtaining ASE from such speciality fibers. Also, stability studies of emission from colloidal QD core fibers have also not been reported.

Herewith, we report anisotropic CdSe/CdS core/shell QDs containing optical fibers for fiber laser applications. The development of such fibers has been done in two steps. Firstly, anisotropic (pod shaped) CdSe/CdS QDs have been fabricated and their morphology optimized for obtaining high optical gain. As fabricated pod-shaped nanocrystals have been investigated in detail using Transient absorption (TA) spectroscopy at 370 and 555 nm excitation wavelength. In the second step, these QDs are used as active gain media in optical fibres. Hollow borosilicate capillaries were filled with the anisotropic morphology QDs dispersed in toluene and investigated for their optical characteristics. Further, ASE has been successfully achieved at RT in the speciality QD fibers by varying QD concentration in the fiber and CW pump power.  Stability of emission from these QD fibers has also been investigated.

## Results

**Fig. 1: Structural and optical properties of pod-shaped CdSe/CdS**

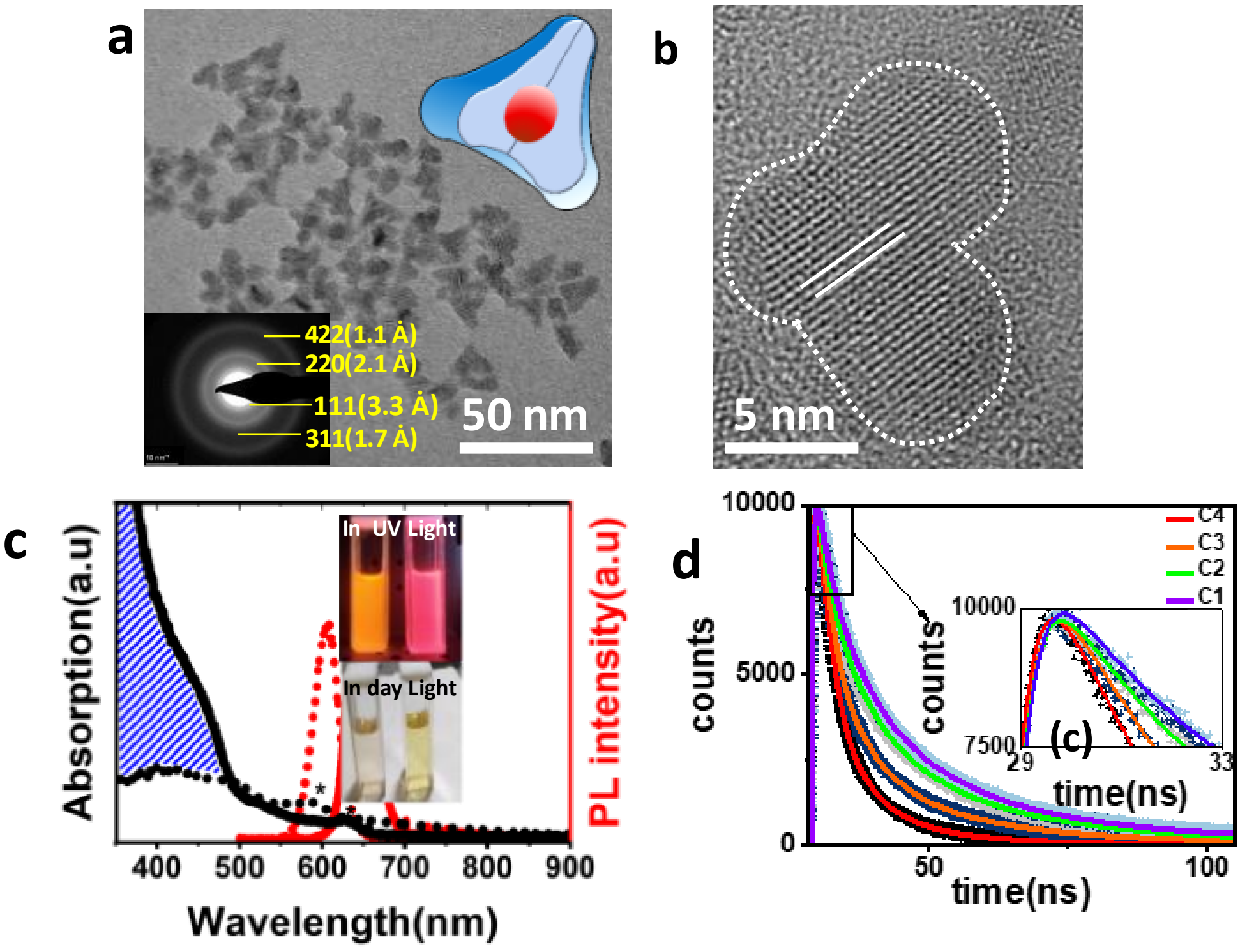

TEM images of **a** CdSe/CdS Core/shell pod shaped QDs. Inset: SAED pattern of QDs along with schematic of the observed morphology **b** HRTEM image of CdSe/CdS QDs **c** emission spectra (red line) and absorption spectra (black line) of CdSe (dotted line) and CdSe/CdS pod shaped QDs (continuous line). Inset shows CdSe QDs (on left) and CdSe/CdS (on right) QDs in day light and under UV light. **d** Fluorescence decay trace for various concentrations of QDs. Inset: shows lifetime increase with a decrease in concentration.

As prepared CdSe and CdSe/CdS core shell QDs were investigated for their structural characteristics using TEM and HRTEM. CdSe QDs are observed to be spherical in shape (Fig S1a) with average particle size of 3 nm (inset Fig S1a). d values calculated from interplanar distances reveal [100], [110], and [002] planes of hexagonal phase of CdSe. When high concentration of CdS monomers were injected at high temperatures, CdS rapidly deposits over CdSe QDs which act as seeds for CdSe/CdS core shell QDs formation. After 15 minutes of reaction, the spherical QDs transform into a pyramid shape of size 7 nm(Fig S1c,d) , which further evolves to pod-shaped particles with average size of ~ 12 nm (Fig 1a). The SAED pattern of the pod shaped structure reveals that the CdS shell formed has hexagonal phase and confirms the presence of [111], [220], [311] and [422] planes of CdS (Inset Fig 1a). HRTEM images further support the pod morphology of the QDs (Fig 1b).

Under UV illumination CdSe QDs emit orange and CdSe/CdS pod-shaped QDs emit red light (Inset Fig 1c). Fig 1c depicts the typical absorption and emission spectrum of CdSe and CdSe/CdS QDs. In the absorption spectra, excitonic absorption at 590 nm and 628 nm can be observed for both CdSe and CdSe/CdS core shell structures. CdSe/CdS QDs show higher absorption for wavelength < 480 nm as compared to CdSe QDs. The CdS shell, makes up the majority of volume fraction of the QDs, and thus a strong absorption due to the shell is highly anticipated[46-47]. Further, the CdSe QDs emission spectra reveals emission maxima at ~ 607 nm which shifts to ~ 645 nm for CdSe/CdS core shell structures. The redshift of the emission maxima can be attributed to the exciton leakage into the shell due to the small offset (0.2 eV) of the conduction band between CdSe and CdS. Low FWHM of emission peaks ~ 34 nm for CdSe and ~ 36 nm for CdSe/CdS QDs also suggests that the monodispersity of the QDs has been well maintained after shelling. The effectiveness of cladding is further confirmed by the fact that the emission spectra is red shifted as compared to CdSe QDs and no extra emission peak from CdS QDs (which emits at lower wavelengths[48]) is observed.

Further, dynamics of radiative exciton recombination are investigated by using time-resolved PL spectroscopy (Fig 1d). QD solution concentration (16, 5.3, 1.8, 0.7 nmol/ml,) was altered while the pulse source energy remained constant (4.8 µW at 1 MHz). Number of excitons generated is given by $\langle N_0 \rangle = j_p \sigma_0$ [34], where $j_p$ is the pump photon fluence (number of photons per unit area per pulse), and $\sigma_0$ is the QD absorption cross section. Pump photon fluence ($j_p$) was kept constant and effective exciton generation was controlled by varying the concentration of the QDs. To fit the decay plot, a tri-exponential curve of the type:

$$\mathrm{I(t)} = \mathrm{I(0)} \sum_{i=0}^{3} A_i e^{-t/\tau_i} \tag{1}$$

where I (0) and I (t) are the PL intensities at times 0 and t, respectively is employed. $\tau_i$ is the excited state lifetime of each component of PL decay, and $A_i$ is the relative amplitude of that component. The fastest decay component increases from 0.6 to 2 ns with increasing absorption cross section, is attributed to the single exciton radiative decay ($\tau_1$). Whereas the moderate decay ($\tau_2$) component is attributed to the Forster resonance energy transfer assisted decay, which increases from 4.41 to 6.89 ns with an increase in absorption cross-section or decrease in concentration. The slowest decay rate ($\tau_3$) has been attributed to the trap states (table S1)

**Fig. 2: Time-resolved TA spectra**

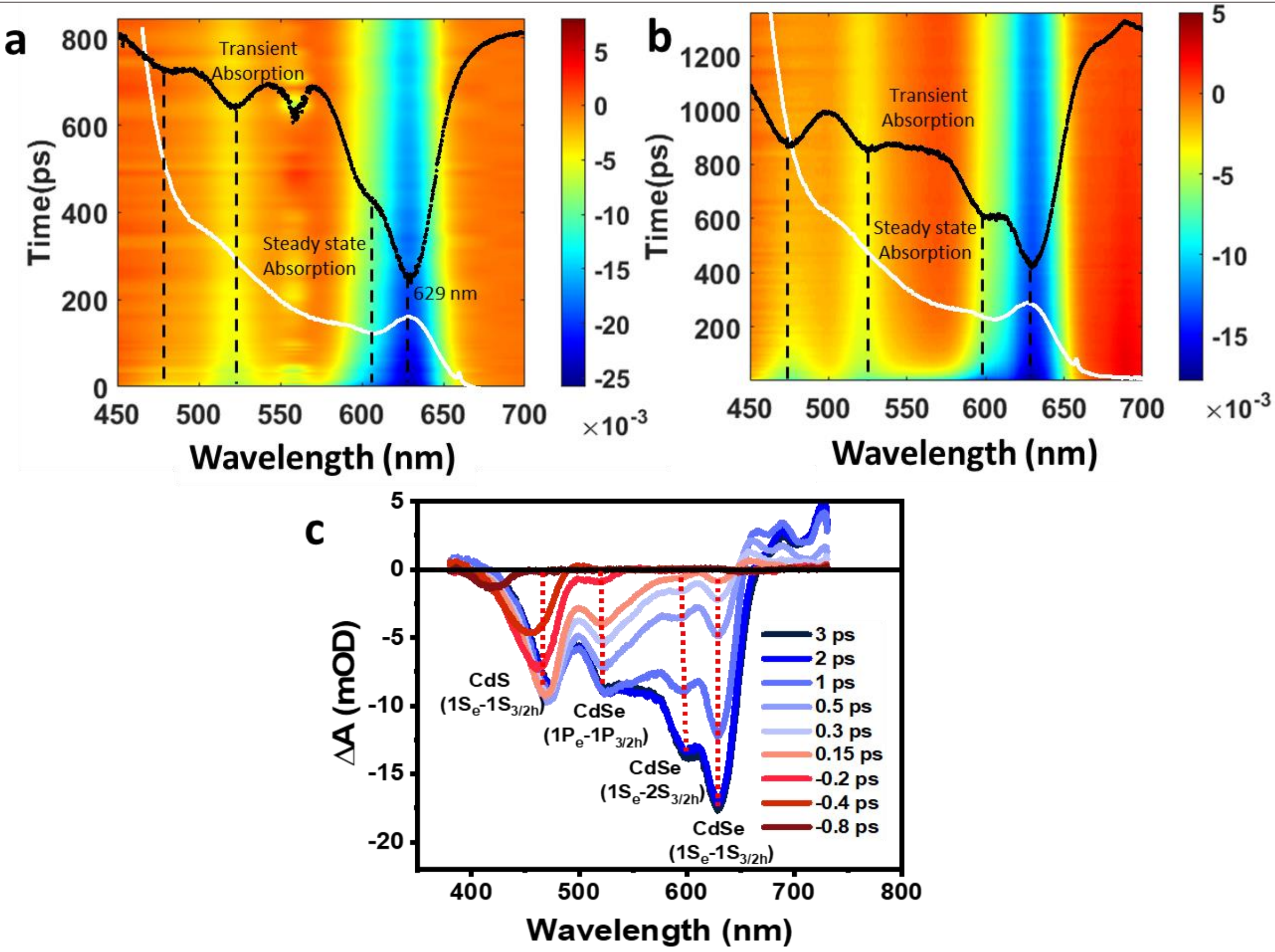

Pseudo color 2D contour plot of Time-resolved TA spectra at **a** 370 nm and **b** 555 nm excitation pump wavelength. Steady state absorption spectra (white) and bleach spectra (black) are also shown at each wavelength. **c** Bleach spectra at various time intervals between pump and probe when excited with 370 nm.

To further explore the transition processes in the pod shaped CdSe/CdS core-clad structures, transient absorption (TA) spectroscopy is used. TA spectra for two excitation wavelengths, 370 nm and 555 nm have been investigated in detail (Fig 2a-2b). From the absorption spectra it is observed that the QDs are highly absorbing under ~ 628 nm wavelength. Therefore, the sample is excited with 370 nm and 555 nm to produce TA signals (Fig 2a). When excited with either 370 nm or 555 nm, the transient bleach signals of the first, second, and third exciton transitions for core of the CdSe/CdS remains constant and are shown by negative amplitude at 629 nm, 598 nm, and 523 nm, respectively (Fig 2a-2b).

**Fig. 3: Gain characteristics of the QDs with fs pumping**

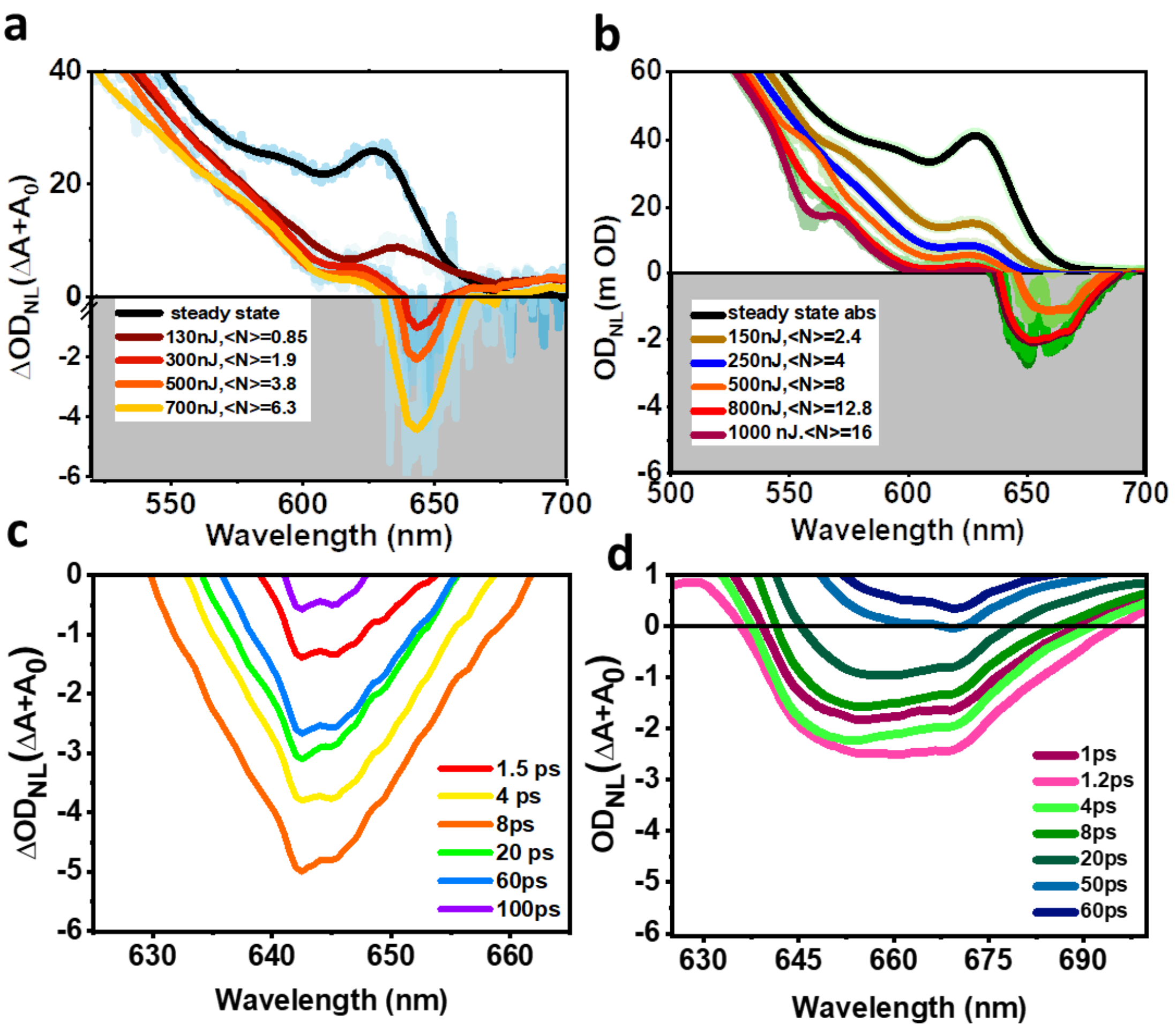

Excited state absorption ($A_{ex}$) obtained from $\Delta A+A_0= A_{ex}$ at different excitation density (pump power), <N> at 5ps delay time corresponding to pump wavelength **a** 370 nm **b** 555nm; Optical gain lifetime corresponding to <N>=16; using pump wavelength as **c** 370 nm, at 700 nJ/Pulse **d** 555nm, at 1000 nJ/Pulse

Bleach signal at 474 nm is due to CdS shell. Absorption peaks in the steady-state absorption spectrum (white fig 2a) are a good match for the first two bleaching bands; the next two transitions, however, were not resolvable (in steady state absorption spectra) but are still clearly evident in the TA spectrum. Bleaching has been attributed to the transitions $1S_e$-$1S_{h3/2}$ (1S), $1S_e$-$2S_{h3/2}$ (2S), and $1P_e$-$1P_{h3/2}$ (1P) in CdSe and $1S_e$-$1S_{h3/2}$ in CdS. Further, bleach spectra at various time intervals between pump and probe have been analysed (Fig 2c). Since excitation wavelength (370 nm) is lesser than CdS absorption edge wavelength, early bleaching at 463 nm due to injection of hot charge carriers take place in CdS (Fig 2c). Bleaching increases as the time delay is changed from -0.8 ps to 0.5 ps. At 0.5 ps, the band reaches its saturation due to complete state filling. Additionally, the bleaching peak at 463 nm shifts to 473 nm with the change in delay time from -0.8 ps to 0.5 ps. This red shift in the bleaching peak indicates a change in the transition levels caused by biexciton interaction. Further, the 1S and 2S CdSe band bleaching appeared at 0.15 ps which continued to develop upto 2 ps, whereas the 1P band has been detected from -0.2 ps which saturated in 1 ps. Time delay dependent bleach spectra proves that hot carriers filled the higher energy excited states in CdS band and 1P band. Later, by Pauli's exclusion principle, the charge carriers relaxed to the lower energy state after completely filling higher energy states. This is the reason for the continuous rise of 2S bands up to 2 ps, whereas the CdS and 1P were saturated in 0.5 ps and 1 ps of time, respectively. It is evident that core state filling did not occur until the CdS band was completely full. Additionally, the CdS bleaching band has long lifetimes (Fig 2b), which is primarily due to easily delocalized electrons in these quasi type-II band aligned nanostructures. There is also a chance of exciton recombination in the CdS band due to the long life of electrons in this band, however, due to the lack of holes, most of the electrons make their way to the core. Additionally, a photo-induced $1S_e$-$1S_{3/2h}$ band can also be observed at ~ 660 nm.

In TA spectrum, ΔA is obtained, which is absorption after excitation of the sample in ground state $\Delta A = A_{ex} - A_0$ ($A_{ex}$ = absorption spectrum when the sample is in an excited state, $A_0$ = steady state absorption). $A_0$ is recorded without changing the concentration of the solution so that the same absorption cross-section is maintained. Net optical gain, which implies that the photo-excited sample will amplify rather than reduce the probe intensity, is identified by negative values of non-linear absorbance $A_{ex} = A_0 + \Delta A$. Fig 3a and 3b shows the non-linear absorbance $A_{ex}$, when excited with 370 nm and 555 nm respectively for 5 ps delay time. As it can be seen, in both cases the optical gain build-up is significant due to inversion asymmetry break up at anisotropic morphologies. With 370 nm excitation, optical transparency (zero $\Delta A+A_o$) is obtained at 130 nJ/pulse for $\langle N \rangle = 0.85$, whereas optical transparency with 555 nm excitation is obtained at 250 nJ for $\langle N \rangle = 4.00$. Evolution of optical gain spectrum with time is demonstrated in fig 3c and 3d. When excited with 370 nm, gain window is attained at a wavelength greater than band-edge absorption feature ( > 628 nm ). The gain window extends from 636 to 657 nm at 1.5 ps and from 640 nm to 647 nm at 100 ps. When excited with 555 nm, the gain window is found to be between 635 nm to 695 nm at 1.2 ps and 667 nm to 672 at 50 ps.

In order to quantify the state filling process and exciton formation in pod-shaped QDs, TA kinetics at lower time delay (between pump and probe) has been studied. Fig. 4 depicts the transition time or occupancy of electron hole pairs in the respective bands when pumped with 370 nm excitation wavelength. The CdS band, which starts at -0.8 ps, has an early 1S state transition and by 0.5 ps reached its saturation. In contrast, the 1P transition in CdSe reached its saturation after 0.38 ps of 1S CdS band filling (difference between filling saturation of 1P and CdS 1S transition) , suggesting that the electron occupancy period in the CdS band is 0.38 ps. 2S and 1S transitions occurred simultaneously due to the fact that they both share the same $1S_e$ electron band. Boltzmann sigmoidal function fitting has been used to determine the overall transition and bleach growth time.

$$A_2 + \frac{A_1 - A_2}{1 + e^{(t-ts)/\zeta r}} \qquad (2)$$

**Fig. 4 : Multiexciton and bi-exciton dynamics**

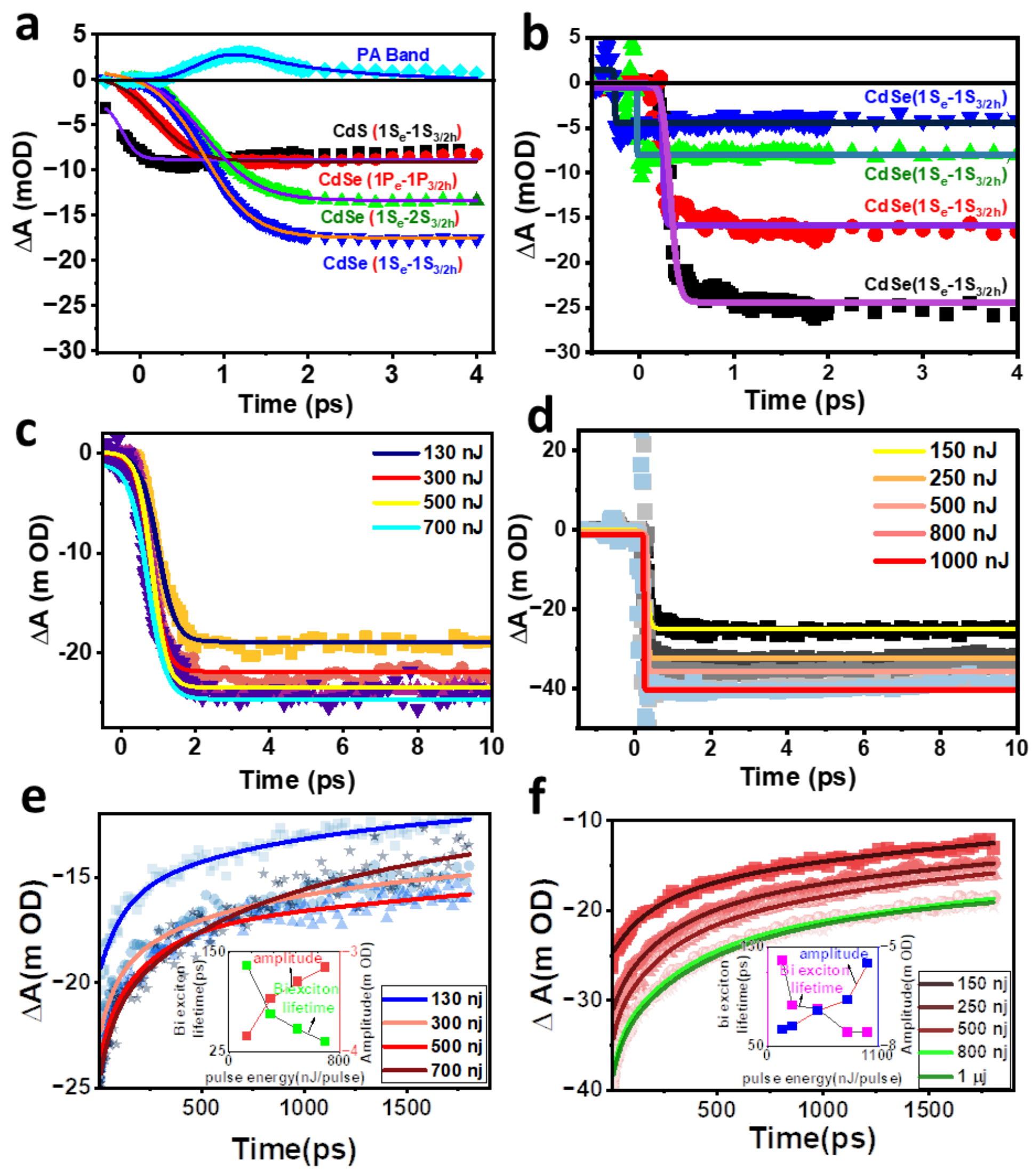

Early time TA signals of different transitions of the CdSe/CdS QDs excited with **a** 370 nm and **b** 555 nm wavelengths, at fixed pump power, 130nJ ; **c**, **d** 1S TA signal of the core when pumped with different energy at lower time intervals between pump and probe; Bleaching decay dynamics with pump power for **e** 370 nm and **f** 555 nm. Inset shows the fitted biexciton lifetime and its corresponding amplitude with respect to pump power.

where $A_1$ is the initial absorbance value, $A_2$ is the final absorbance value, $t_s$ indicates the time center, and $\tau_r$ represents the time constant for the rising process. The rise time constants for CdS 1S and CdSe 1P, 2S, and 1S transition are 0.09 ps, 0.23 ps, 0.31 ps and 0.29 ps. A photo-induced band from 1S state is also present. The rising process gets over in 1 ps and dies down exponentially due to localization and relaxation. Kinetics of TA demonstrates no induced absorption when excited with 555 nm Fig S2) excitation at lower time intervals between pump and probe. In comparison to 370 nm, the transition associated with each band reached saturation in shorter durations at 555 nm excitation (Fig 4b). For CdS 1S and CdSe 1P, 2S, and 1S, the rise times for the transitions to achieve saturation were calculated to be 0.004, 0.001, 0.01, and 0.04 ps respectively. Additionally, saturation in 1S transition states indicates inter band carrier relaxation has reached equilibrium and that there is no band edge recombination. The electron population in the CdS band has a lifespan of 0.23 ps before relaxing to the 1P state. Electron occupation lifespan in the 1P state is around 0.30 ps. 2S and 1S band transitions begin concurrently, indicating electron relaxation from the 1P state to the 1S state.

TA kinetics of the 1S state at an early time with variable pump power are shown in Fig. 4c. With the Boltzmann sigmoidal function fitted to the state filling process, it was discovered that, regardless of pump power, saturation is reached in 0.2 ps. It is evident that when pump power increases, multiexciton generation will also increase, which will lead to an increase in intrinsic auger recombination. The saturation of ΔA has been identified to be constant up to 10 ps for the highest pump power. This saturation in ΔA also indicates that there is no electron-hole recombination and also occupancy of electrons in the $1S_e$ state and hole occupancy in $1S_{3/2}$ is stable. The amount of time for which ΔA stays saturated is a direct measure of bi exciton lifetime. The lifetime for ΔA saturation increases with the reduction of pump power, due to reduction in auger recombination. The corresponding decay kinetics of ΔA has been plotted in fig 4e. The bleaching fits well with a summation of 2 exponential decays. Shorter lifetime ($\tau_1$) is attributed to biexciton decay, for which the probability to observe stimulated emission is the highest, and the longer one ($\tau_2$>1 ns) is attributed to single exciton lifetime. The biexciton lifetime has been of major interest since the generation of biexciton is a key requirement to obtain optical gain. The biexciton lifetime (fitted) and amplitude of the fitting component have been shown with respect to pulse energy in fig 4e inset. Further, it can be seen that with an increase in pulse energy the amplitude increases since higher pulse energy would create more number of bi-excitons. However, the lifetime of the biexciton decreases with pulse energy because high energy would create more multi excitons and thus auger recombination probability will also increase leading to a decrease in biexciton lifetime due to enhanced auger recombination.

Fig 4d shows early-time TA kinetics when excited with 555 nm with varying pump power. The state-filling process attained saturation quickly compared to when excited with 370 nm. The rise time is found to be 0.04 ps, which remains almost constant with variation in pump power. It has to be noted that the rising time of 1S state is shorter as compared to 370 nm excitation because 555 nm aids transition-related to $1S_e$ state since its corresponding wavelength is closer to this excitation wavelength. As it has been shown that 555 nm also creates bleaching in higher energy states, ΔA up to 10 ps has been attributed to occupancy of states by multiexcitons. The bleaching decay of the 1S state with pump power has been shown in fig 4f. A summation of two exponential decay has been fitted to the data to extract biexciton and single exciton dynamics. The fitted biexciton lifetime and corresponding amplitude with pump power have been plotted in the inset. A Similar decrease in biexciton lifetime and increase in amplitude can be seen.

Depletion of ground state carriers to excited states is referred to as "bleaching of the ground state." Stimulated emission, which is Stokes-shifted because of the coulombic interaction of the biexciton, in relation to the bleach signal and frequently still overlaps with it, follows the quantum dot's photoluminescence spectrum. This is a quantum dot lasing effect (coherent emission). This transmitted signal typically causes erroneous negative absorbance peaks in the final spectra because it cannot be separated from the absorption signal. To distinguish between ground-state bleach and stimulated emission, the spectra has been deconvolved. In Fig 5, ΔA for different pump power at a fixed delay time of 5 ps has been deconvoluted with 6 peaks. The selection of peaks for various bands has been done by strictly following the steady-state absorption. The stimulated emission from an ensemble of quantum dots has been approximated as gaussian due to gaussian size distribution of the quantum dots. The stimulated emission band spans the 2S transition since 1S and 2S shares a common $1S_e$ electron occupancy level. The peak position has been approximated from the previously observed non-linear $A_{ex}$ spectrum. It can be seen that a significant signal by stimulated emission overlaps with 1S absorption, which also suggests that overall stimulated emission observed from the system will always be redshifted. The area under the stimulated emission peak that did not overlap with an absorption peak represents the strength of the stimulated emission. The absolute area of this region was found to increase from 0.06 to 0.28 with the increase in pump power from 150 to 1000 nJ (covered in black). The spectrum resembles the negative region of the $A_{ex}$ (Fig 5f), which corresponds to gain or stimulated emission from the QDs.

**Fig 5: De convulation of TA spectrum to extract strength of stimulated emission**

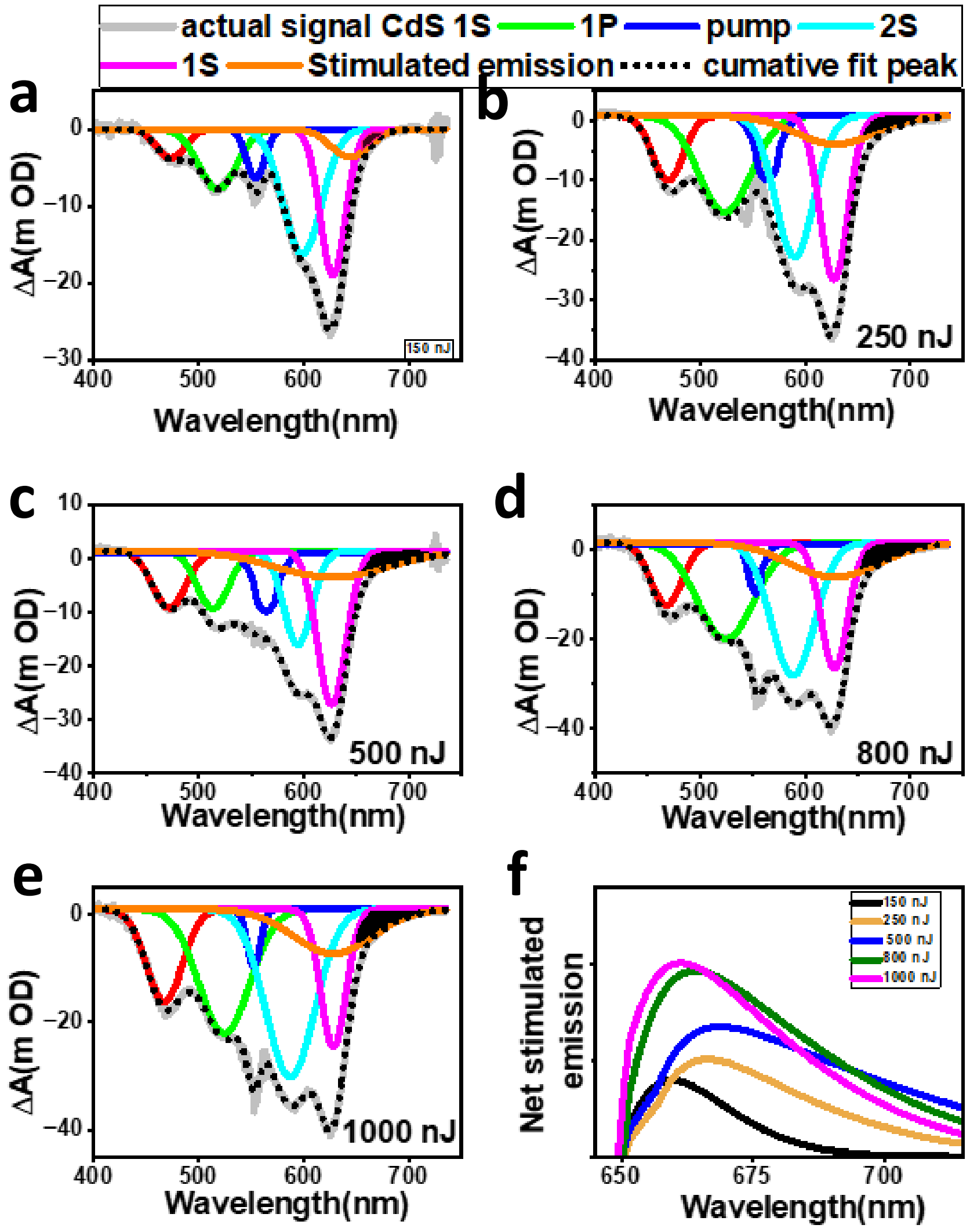

Deconvoluted ΔA for different pump powers **a** 150 nJ **b** 250 nJ **c** 500 nJ **d** 800 nJ **e** 1000 nJ. **f** Stimulated emission signal extracted at each pump power.

After obtaining gain in CdSe/CdS QDs, they were entrained in borosilicate hollow capillaries in order to check the effect of waveguide on optical characteristics of QDs. First requisite for confinement of QD emission in the waveguide is higher effective refractive index of the core. A proper investigation of the refractive index (n) of QD solution is carried out in order to check the waveguiding properties of the QD doped fibre. From the refractive index of the curve [49], it can be observed that n is almost constant (n = 1.51) from 450 to 750 nm (Fig S3), with a negligeable variation of 1%. Since 532 nm pump is being used to excite QDs in the core of the fiber and emission is at 650 nm, therefore it is safe to conclude that n is constant in the region of interest. Further, refractive index of capillary clad borosilicate glass at 650 nm is 1.51, ensuring total internal reflection in the core.

**Fig. 6: QD filled fiber**

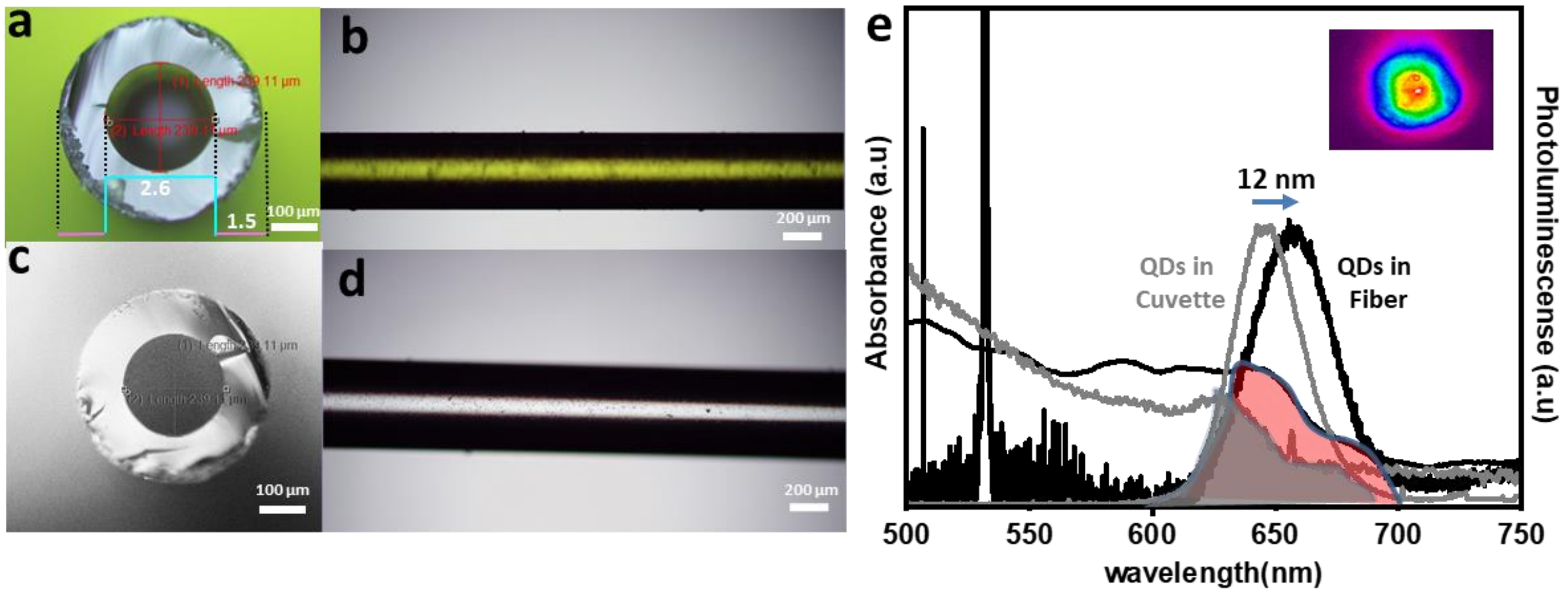

**a** Vertical and **b** Horizontal cross section of a hollow capillary; **c** Vertical and **d** Horizontal cross section of a capillary filled with QDs (core diameter = 240 μm, length = 20 cm) **e** Absorbance and PL of QD solution in a cuvette (grey) and QD doped fibre (black). Inset: Near field image of the core shows the multimode behaviour of QD emission in the core

The QD-filled fibre was first observed under an optical microscope, to confirm the uniform filling of QD solution inside the fibre. A longitudinal and transverse optical image of the fibre before and after filling with QD solution was taken (Fig 6a-6d). To obtain the absorption spectra of the QD-filled fibre, first the fibre was filled with toluene and then a white light source was coupled with a 20X objective lens, the spectra were collected using OSA. Later the same fibre was filled with QD-toluene solution and spectra were collected after coupling with the white light source. For obtaining the PL emission the same fibre was coupled with 532 nm CW laser and emission was recorded in OSA. Fig S4 depicts the uniformly filled QD fiber emitting orange when coupled with a 532 nm green laser source.

Both absorption and emission spectra are red-shifted in QD filled fibre as compared to QDs in a cuvette (Fig 6e). The red shift can be attributed to a phenomenon called reabsorption-emission due to overlap between emission and absorption spectra as observed in Fig 6e, which weakens the signal from smaller QDs. Another reason for such a redshift could be interaction between the borosilicate and QD in toluene interface. Further, emission characteristics of QD filled fibers were studied by varying the fiber length from 35 to 120 cm (60 um core radius) at 10 mW of pump powers. It is observed that the intensity of emission decreases and peak position shifts to a longer wavelength (Inset Fig S5) with the increase in length of the fibre.

**Fig. 7 ASE build up with increase in concentration**

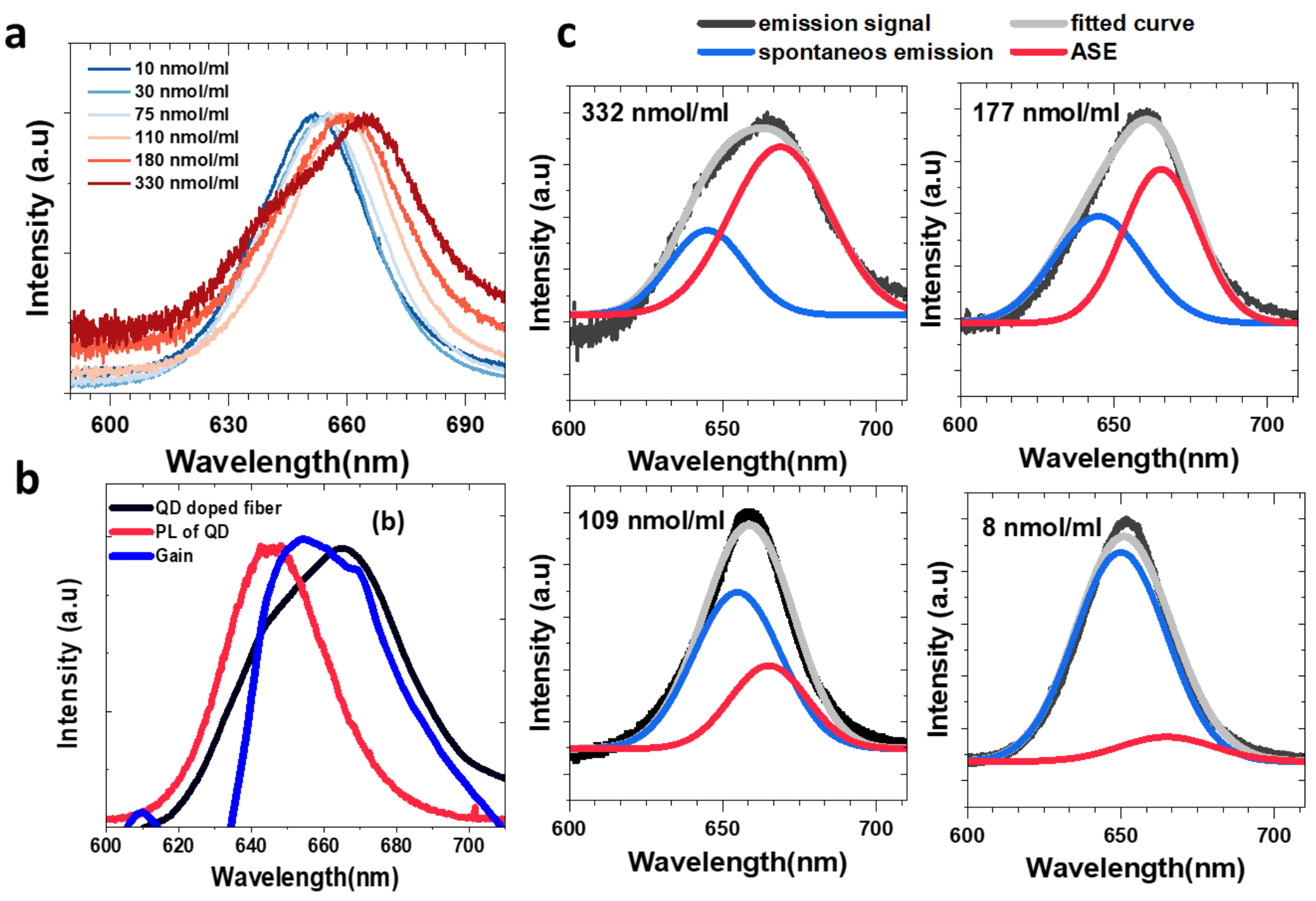

**a** Variation in emission spectra with concentration of CdSe/CdS QDs in hollow glass capillaries of diameter (275μm) and length (70 cm) excited with fixed pump power (50 mW), **b** Comparison of PL spectra of QDs in a cuvette, gain obtained from transient absorption spectrum (pump wavelength 555nm) and PL of QD filled fiber with the highest concentration of QD solution at pump power 400 mW. **c** Peak fitted for emission spectra obtained from QD-filled fiber with different concentrations, to have a quantitative idea of spontaneous emission and ASE

Evolution of emission spectrum by varying core diameter of QD filled fibers has also been studied (Fig S6). Fibre with core diameters of 110, 240, 275, 370 and 535 um with equal lengths (35 cm) have been filled with the same concentration of QD solution and pumped with 10 mW - CW green laser. It can be seen that the intensity of transmitted emission spectra increases initially with increase in core diameter from 110 to 275 um. With further increase in the core diameter the emission intensity drops down. Initially on increasing the fibre diameter the pump power will be distributed over larger area and a greater number of QDs can be excited, leading to an increase in intensity. By further increasing the diameter, the pump power will have more spread due to which it will be totally absorbed on short propagation and the rest of the length will then be a source of losses due to the liquid medium leading to a decrease in emission intensity.

It has been shown that through fs pumping the QDs dispersed in toluene show optical gain for appropriate pump powers (Fig 3d). The emission taken from QD fibers, filled with different concentrations of QD solution is shown in Fig 7a. It is observed that with increase in concentration the emission maxima red shifts. Further, at QD concentration ~ 330 nmol/ml a separate peak shows up. The red shift in emission can be attributed to the growth of ASE. To further confirm the ASE, a comparison of the emission spectra from QD fiber, QDs in a cuvette and gain obtained from fs pumping of the QDs has been illustrated in Fig 7 b. It is clear that the assigned ASE signal overlaps with the gain obtained from fs pumping. Also, it is evident that the emission spectra from QD fibers is a superposition of two phenomena: spontaneous emission from QDs (645 nm) and amplified spontaneous emission (ASE, 669 nm). This is also illustrated by deconvoluting the emission spectra (Fig 8c) using a gaussian profile. The peak position of the gaussian corresponding to Spontaneous emission has been fixed from the PL peak of the QDs in solution (645 nm), whereas the ASE gaussian peak has been chosen from the gain spectrum that has been obtained from femtosecond pumping. As the concentration is increased from 10 nmol/ml to 330 nmol/ml i.e., increase in packing fraction, the contribution from spontaneous emission decreases while that from ASE increases.

In order to investigate ASE further, high concentration of QDs were exposed to different pump powers. It was anticipated that at higher concentration, due to enhanced QD-QD interaction, emission from a QD will trigger a chain of downfall of excitons to the ground state in which the radiation from one QD stimulates another in succession resulting in stimulated emission. Also, high pump power would ensure minimal reabsorption and QDs are excited through pump power only. Emission from 330 nmol/ml has been obtained by exposing the QD fibers to pump powers varying from 10 mW to 400 mW. The emission spectrum taken for the fiber filled with 330 nmol/ml concentration of QDs is shown in Fig 8a. The spontaneous emission peak from the QDs and ASE peak has been identified. It is clear that with the increase in pump power, the ASE peak gets intensified. The peak power corresponding to these two phenomena and their power difference has been shown in fig 8b. Following the subtraction of the background resulting from the "saturated" spontaneous emission, this new band shows a well-defined threshold (~520 $Wcm^{-2}$) and a clear linear growth; both of these observations are signatures of the ASE.

**Fig. 8: Demonstration of ASE**

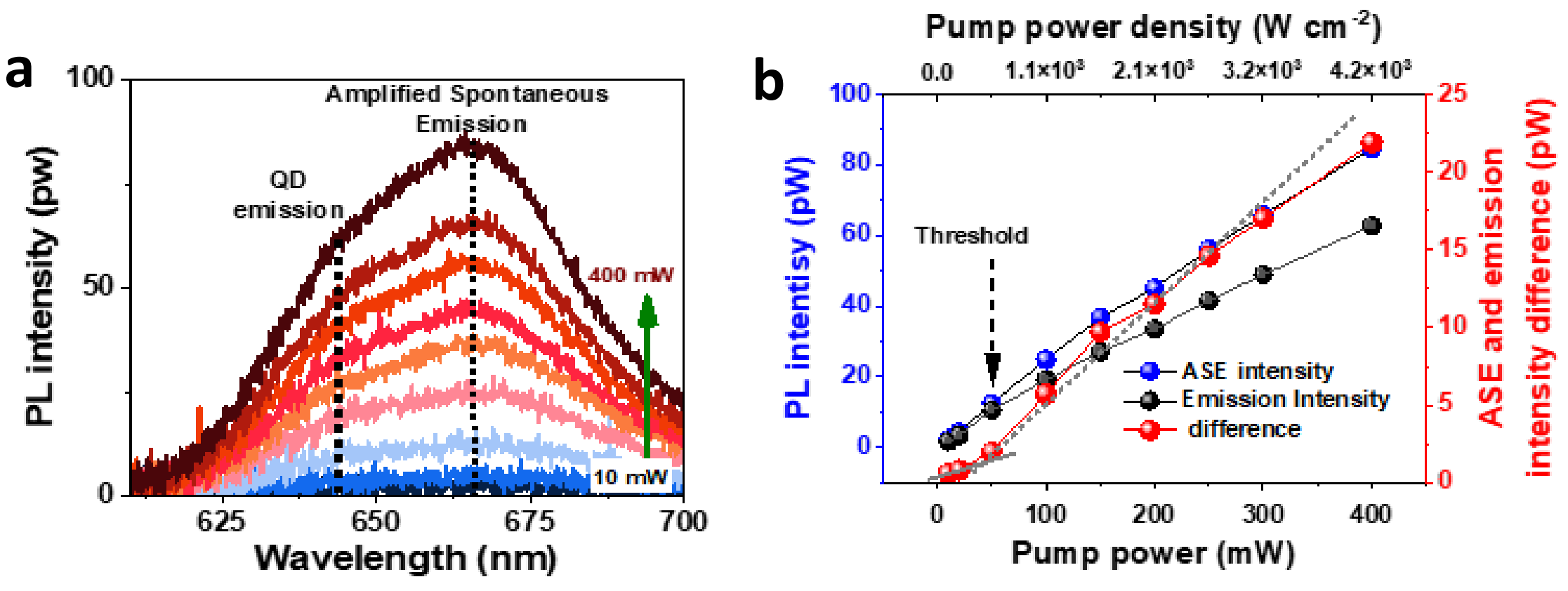

**a** Development of ASE band at ~ 669 nm is observed as a function of CW-pump power. Pump power is varied from 10 mW to 400 mW. **b** The ASE band is seen to grow linearly. A definite threshold of 50 mW(520 Wcm$^{-2}$) is shown when the strong spontaneous PL that is present in the ASE spectrum's background is subtracted.

In order to obtain ASE, high powers (0.4 kW) have been used which may lead to degradation of the QDs through photobleaching. To understand photobleaching and get an estimate of maximum power which these QDs can sustain in fiber without degradation, a highly concentrated (330 nmol/ml) and a highly diluted (5 nmol/ml) QD solution drop casted on a glass slide were kept under a confocal microscope and exposed to a green 532 nm-CW laser with excitation power of 200 mW. The beam has been focused with 60X scope(spot size = 0.6 μm), making the intensity to be 100 kW cm$^{-2}$. The intensity was collected with CCD camera with time(Fig S7), later the image's overall intensity was calculated over an area, the intensity drops down with time and the intensity vs time curve was fitted with a decay curve (Fig 9a). With diluted solution, time taken (t) for intensity to reach e$^{-1}$ from initial intensity was 13 secs whereas, when a concentrated solution was irradiated under the same condition, t was 138 secs. This suggests that with increase in no. of QDs the heat produced by laser gets more distributed hence the occurrence of photobleaching gets minimized.

To test the stability of QD fibre under high power, a 70 cm QD fibre with a uniform core diameter (110 μm) was constantly irradiated with 400 mW laser for 30 mins. Emission spectra recorded after every five minutes shows that maximum intensity appeared at time = 0 secs. On further exposure to such high pump power, intensity dropped by 25 % in 5 mins and then the emission intensity was constant. This indicates that, QDs present at the beginning of the fibre receive most of the power which get photobleached faster, whereas rest of the QDs present over the length of the fibre remains unaffected and hence emission remains constant after 5 mins.

Thanks to large surface to volume ratio of the fibre which leads to natural cooling and hence the emission signal from the fibre remains unaffected with time.

**Fig. 9: Emission stability of the QD specialty fiber**

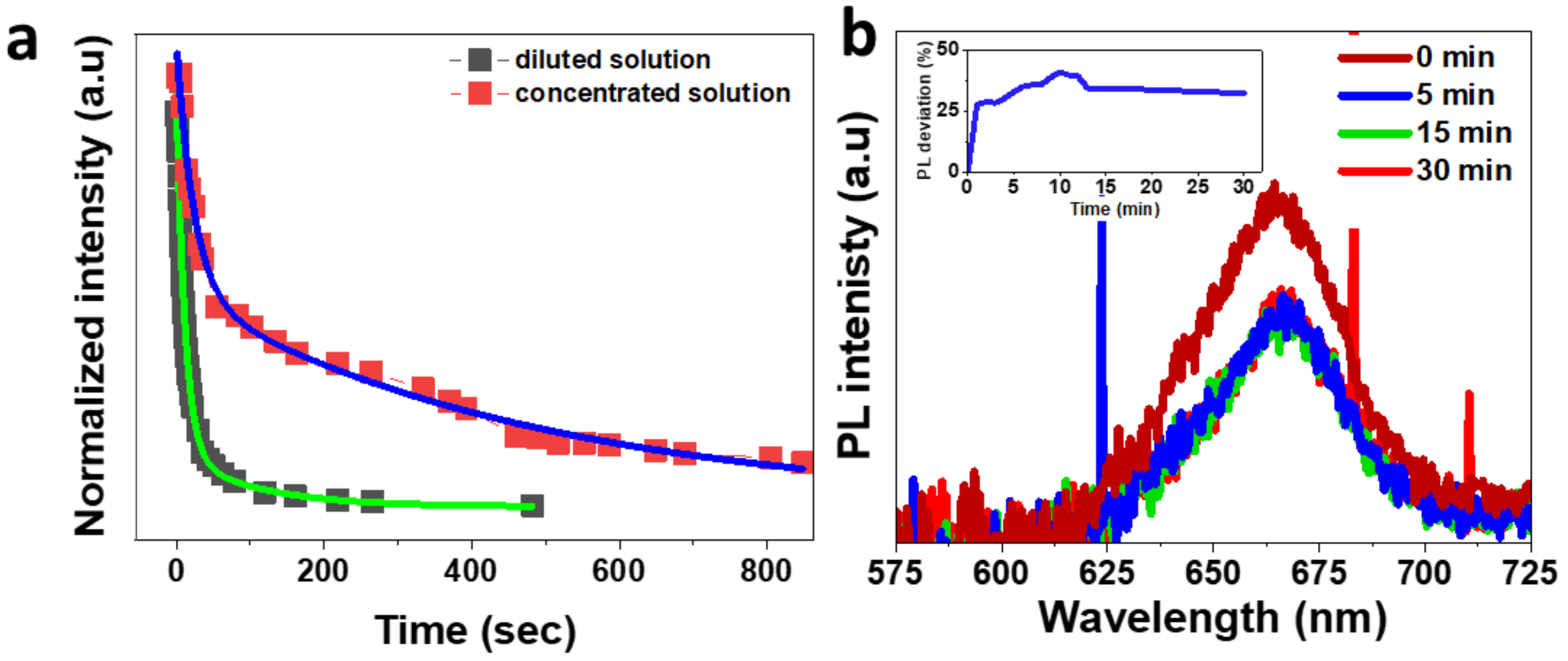

(a) Variation of normalized PL intensity with time for diluted and concentrated CdSe/CdS QDs (b) Emission spectrum of QD-filled fiber with time, under 400 mW power exposure. Inset shows that PL peak intensity variation with time. Emission intensity stays constant for 5 to 30 mins of exposure.

## Discussion

In conclusion we have synthesized asymmetric shaped pod shaped CdSe/CdS QDs which shows efficient optical gain through fs pumping. The same QDs has been used to fabricate specialty fiber. The emission properties of the QDs has been characterized. Later , ASE build up in this fiber has been shown with increase in concentration of the QDs i.e., increase in packing fraction under CW excitation. Also, ASE has been demonstrated with increase in pump power with clear indication of threshold power of 520 W cm-2. These results suggest the feasibility of actual fiber laser with tunable emission is possible with implementation of proper cavity.

## Methods

### Synthesis of CdSe/CdS quantum dots

The CdSe/CdS core shell QDs are fabricated by a colloidal synthesis technique. 0.2 mmol CdO, 8 mL ODE, 8.2 ml of OlAm and 0.3 mL OA were placed in a 250 mL 3-neck flask and degassed for 10 min. The solution was further heated to 270 °C under argon-gas flow. At this temperature, 2 mmol TOP-Se in 2 ml of ODE suspension was quickly injected into the flask, and the reaction was allowed to proceed for 5 min. Finally, the reaction mixture was cooled to room temperature, and an extraction procedure by centrifugation using ethanol for 10 minutes was performed to purify the QDs from side products and unreacted precursors. For shelling, the QDs were redispersed in 4 ml of ODE in a three-neck flask and degassed for 10 minutes and the temperature was set to 200 °C. At this temperature 1 mmol of sulphur dissolved in TOP and ODE was injected into the three-neck flask. After 60 mins, 2 mmol of previously prepared Cd-oleate dissolved in ODE was injected. The temperature was further maintained for 60 mins and allowed to cool down to room temperature before the extraction procedure. After purification, the QDs were dispersed in toluene and used for further characterization.

### Fabrication of CdSe/CdS filled optical fibre

The QDs dispersed in toluene at different concentrations were filled in inhouse fabricated borosilicate capillaries with different core diameters ranging from 110 µm - 537 µm. Filling of the fiber was done by submerging one end of the hollow capillary in QD solution and by pressure pulling the QD solution from the other end.

## Characterization

UV–vis absorption spectra was collected on S 600 Analytik-jena. Photoluminescence (PL) spectra of QDs were recorded using ocean optics USB200. Transmission and high-resolution electron microscopy (TEM, HR-TEM) images were obtained on TEM-TITAN Themis. Lifetime was calculated using Horiba Jobin Yuvon Fluorocube -01-NL Fluorescence Lifetime System. For characterization of QD filled fibre CW-532 nm (GEM532, Laser Quantum), Optical Spectrum Analyzer (OSA) (AQ6374, Yokogawa) and supercontinuum source (LEUKOS) were used.

## Transient Absorption Spectroscopy

A Ti: sapphire regenerative amplifier (Spitfire, Spectra-Physics) with an output of about 2 mJ/pulse, pulse width 100 fs, and a centre at 790 nm was used to amplify seed pulses of 100 fs that were produced by a Ti: sapphire oscillator (Tsumani, Spectra Physics) at a repetition rate of 1 KHz. A beam splitter was used to divide the enhanced 1 mJ/energy pulses into two equal parts (95:5). TOPAS-C(Spectra-Physics) was pumped with 95% of the available energy to create an actinic pump with a 370 nm centre. To create a broadband super-continuum white light (380-720 nm) probe, the remaining 5% of the light is focused on a 3 mm $CaF_2$ crystal. At Magic-angle (54.70°), the pump and probe beams were maintained. Both the beam were spatially overlapped on the moving 1 mm sample cuvette to avoid photodegradation of the sample. A transmitted white light probe and reference beam were selected to enter into the spectrograph (Triax 550, Horiba) and dispersed by 150 mm grooves/mm grating.

.

# Data Availability

Upon reasonable request, the corresponding author will provide the data that back up the study's conclusions.

# References

1. B. Bouzid, "Analysis and review of Erbium doped fiber amplifier," 2013 Saudi International Electronics, Communications and Photonics Conference, 2013, pp. 1-5, doi: 10.1109/SIECPC.2013.6551005.
2. https://www.ipgphotonics.com/en/products/lasers/high-power-cw-fiber-lasers for high power fiber laser product offerings from IPG Photonics, Oxford, MA, USA
3. Materials for optical fiber lasers: A review editors-pick Applied Physics Reviews 5, 041301 (2018); https://doi.org/10.1063/1.5048410 P. D. Dragic, M. Cavillon and J. Ballato
4. E. Desurvire, Erbium-Doped Fiber Amplifiers: Principles and Applications ( John Wiley & Sons, New York, 1994).112.
5. M. J. F. Digonnet, Rare-Earth-Doped Fiber Lasers and Amplifiers, 2nd ed. ( CRC Press, Boca Raton E. Desurvire, Erbium-Doped Fiber Amplifiers: Principles and Applications ( John Wiley & Sons, New York, 1994).112.
6. M. J. F. Digonnet, Rare-Earth-Doped Fiber Lasers and Amplifiers, 2nd ed. ( CRC Press, Boca Raton, FL, 2001)., FL, 2001).
7. G. H. Dieke and H. M. Crosswhite, Appl. Opt. 2, 675 (1963). https://doi.org/10.1364/AO.2.00067589.
8. R. Reisfeld, A. Patra, G. Panczer, and M. Gaft, Opt. Mater. 13, 81 (1999). https://doi.org/10.1016/S0925-3467(99)00015-4
9. M. Hemenway, W. Urbanek, D. Dawson, Z. Chen, L. Bao, M. Kanskar, M. DeVito, D. Kliner, and R. Martinsen, Proc. SPIE 10086, 1008605 (2017). https://doi.org/10.1117/12.225065618.
10. Y. Kasai, S. Sakamoto, Y. Takahashi, K. Katagiri, and A. Sakamoto, Fujikura Tech. Rev. 2016, 42. https://doi.org/10.1049/piee.1966.018919.
11. J. Biesenbach, “ High brightness fiber coupled modular diode laser platform,” in 2012 IEEE Photonics Society Summer Topical Meeting Series (IEEE, 2012), Paper No. MA4.3.

12. Guoping Dong, Haipeng Wang, Guanzhong Chen, Qiwen Pan and ianrong Qiu Quantum dot-doped glasses and fibers: fabrication and optical properties, Front. Mater., 23 February 2015, https://doi.org/10.3389/fmats.2015.00013
13. Jialong Zhao et.al. “Efficient CdSe/CdS Quantum Dot Light-Emitting Diodes Using a Thermally Polymerized Hole Transport Layer” https://doi.org/10.1021/nl052417e
14. W. Cao, C. Xiang, Y. Yang, Q. Chen, L. Chen, X. Yan, and L. Qian, “Highly stable QLEDs with improved hole injection via quantum dot structure tailoring,” Nat. Commun. 9(1), 2608 (2018).
15. H. Zhang, S. Chen, and X. W. Sun, “Efficient Red/Green/Blue Tandem Quantum-Dot Light-Emitting Diodeswith External Quantum Efficiency Exceeding 21,” ACS Nano 12(1), 697–704 (2018).
16. ui Din et al, Recent advances in quantum dots-based biosensors for antibiotics detection,Journal of Pharmaceutical Analysis,Volume 12, Issue 3,2022,https://doi.org/10.1016/j.jpha.2021.08.002.
17. Ling Yun, Yang Qiu, Conghao Yang, Jie Xing, Kehan Yu, Xiangxing Xu, and Wei Wei, "PbS quantum dots as a saturable absorber for ultrafast laser," Photon. Res. 6, 1028-1032 (2018)
18. Angela M. Wagner, Jennifer M. Knipe, Gorka Orive, Nicholas A. Peppas,“Quantum dots in biomedical applications”https://doi.org/10.1016/j.actbio.2019.05.022.
19. heng Cheng, Nengshu Hu, Xiaoyu Cheng,“Experimental realization of a PbSe quantum dot doped fiber amplifier with ultra-bandwidth characteristic”,Optics Communications, https://doi.org/10.1016/j.optcom.2016.08.036.
20. Ali Hreibi, “Semiconductor-doped liquid-core optical fiber” Optics Letters Vol. 36, Issue 9, pp. 1695-1697 (2011) ,https://doi.org/10.1364/OL.36.001695
21. C. Cheng, J. Bo, J. Yan and X. Cheng, "Experimental Realization of a PbSe-Quantum-Dot Doped Fiber Laser," in IEEE Photonics Technology Letters, vol. 25, no. 6, pp. 572-575, March15, 2013, doi: 10.1109/LPT.2013.2243139.

22. Zhang, L., Zhang, Y., Wu, H. et al. Multiparameter-dependent spontaneous emission in PbSe quantum dot-doped liquid-core multi-mode fiber. J Nanopart Res 15, 2000 (2013). https://doi.org/10.1007/s11051-013-2000-z
23. Lei Zhang et al , "Stimulated emission and optical gain in PbSe quantum dot-doped liquid-core optical fiber based on multi-exciton state",Optics Communications,https://doi.org/10.1016/j.optcom.2016.02.048.
24. Lei Zhang, Liping Zhao, and Youjin Zheng, "Enhanced emission from a PbSe/CdSe core/shell quantum dot-doped optical fiber," Opt. Mater. Express 8, 3551-3560 (2018)
25. R. Bahrampour et al "An inhomogeneous theoretical model for analysis of PbSe quantum-dot-doped fiber amplifier" Optics Communications,Volume 282, Issue 22,https://doi.org/10.1016/j.optcom.2009.08.003.
26. Suzanne Bisschop et al " The Impact of Core/Shell Sizes on the Optical Gain Characteristics of CdSe/CdS Quantum Dots" ACS Nano 2018, 12, 9, 9011–9021, https://doi.org/10.1021/acsnano.8b02493
27. Guzelturk, B., Kelestemur, Y., Gungor, K., Yeltik, A., Akgul, M.Z., Wang, Y., Chen, R., Dang, C., Sun, H. and Demir, H.V. (2015), Stable and Low-Threshold Optical Gain in CdSe/CdS Quantum Dots: An All-Colloidal Frequency Up-Converted Laser. Adv. Mater., 27: 2741-2746. https://doi.org/10.1002/adma.201500418
28. Castelli, A., et al., "Core/Shell CdSe/CdS Bone-Shaped Nanocrystals with a Thick and Anisotropic Shell as Optical Emitters". Adv. Optical Mater. 2020, 8, 1901463. https://doi.org/10.1002/adom.201901463
29. R. Brescia, K. Miszta, D. Dorfs, L. Manna, G. Bertoni, "Birth and Growth of Octapod-Shaped Colloidal Nanocrystals Studied by Electron Tomography"J. Phys. Chem. C 2011, 115, 20128.
30. S. Deka, K. Miszta, D. Dorfs, A. Genovese, G. Bertoni, L. Manna," Octapod-Shaped Colloidal Nanocrystals of Cadmium Chalcogenides via "One-Pot" Cation Exchange and Seeded Growth" Nano Lett. 2010, 10, 3770.

31. Xudong Wang," Anisotropic Arm Growth in Unconventional Semiconductor CdSe/CdS Nanotetrapod Synthesis Using Core/Shell CdSe/CdS as Seeds" J. Phys. Chem. C 2019, 123, 31, 19238–19245
32. Joanna Xiuzhu et. Al.,"Quantification of the Photon Absorption, Scattering, and On-Resonance Emission Properties of CdSe/CdS Core/Shell Quantum Dots: Effect of Shell Geometry and Volumes" Anal. Chem. 2020, 92, 7, 5346–5353
33. Klimov, V. I et al. "Quantization of Multiparticle Auger Rates in Semiconductor Quantum Dots" American Association for the Advancement of Science, https://doi.org/10.1126/science.287.5455.1011
34. Li Wang, et al "Quasi-Type II Carrier Distribution in CdSe/CdS Core/Shell Quantum Dots with Type I Band Alignment" J. Phys. Chem. C 2018, 122, 22, 12038–12046
35. Jier Huang, et.al., "Multiple Exciton Dissociation in CdSe Quantum Dots by Ultrafast Electron Transfer to Adsorbed Methylene Blue"J. Am. Chem. Soc. 2010, 132, 13, 4858–4864
36. Haiming Zhu,, et.al., "Controlling Charge Separation and Recombination Rates in CdSe/ZnS Type I Core−Shell Quantum Dots by Shell Thicknesses" J. Am. Chem. Soc. 2010, 132, 42, 15038–15045
37. JCPDS No. 65–2891
38. D. J. Richardson, J. Nilsson, and W. A. Clarkson, "High power fiber lasers: current status and future perspectives [Invited]," J. Opt. Soc. Am. B 27, B63-B92 (2010)
39. Whitworth, G.L., Dalmases, M., Taghipour, N. et al. Solution-processed PbS quantum dot infrared laser with room-temperature tunable emission in the optical telecommunications window. Nat. Photon. 15, 738–742 (2021). https://doi.org/10.1038/s41566-021-00878-9
40. Chhantyal, P., Naskar, S., Birr, T. et al. Low Threshold Room Temperature Amplified Spontaneous Emission in 0D, 1D and 2D Quantum Confined Systems. Sci Rep 8, 3962 (2018). https://doi.org/10.1038/s41598-018-22287-9

41. Mikhail Zamkov et. Al., Quantum Shells Boost the Optical Gain of Lasing Media,ACS Nano 2022 16 (2), 3017-3026, DOI: 10.1021/acsnano.1c10404
42. Hilmi Volkan Demir et. al., Tunable amplified spontaneous emission from core-seeded CdSe/CdS nanorods controlled by exciton–exciton interaction. NanoScale ,Issue 15, 2014 DOI:https://doi.org/10.1039/C4NR01538J
43. Suzanne Bisschop, Pieter Geiregat, Tangi Aubert, and Zeger Hens, The Impact of Core/Shell Sizes on the Optical Gain Characteristics of CdSe/CdS Quantum Dots. ACS Nano 2018, 12, 9, 9011–9021 https://doi.org/10.1021/acsnano.8b02493
44. Fang Yuan, Zhaoxin Wu, Hua Dong, Jun Xi, Kai Xi, Giorgio Divitini, Bo Jiao, Xun Hou, Shufeng Wang, and Qihuang Gong. The Journal of Physical Chemistry C 2017 121 (28), 15318-15325. DOI: 10.1021/acs.jpcc.7b02101
45. Brenner, P., Bar-On, O., Jakoby, M. et al. Continuous wave amplified spontaneous emission in phase-stable lead halide perovskites. Nat Commun 10, 988 (2019). https://doi.org/10.1038/s41467-019-08929-0
46. Synthesis and Micrometer-Scale Assembly of Colloidal CdSe/CdS Nanorods Prepared by a Seeded Growth Approach , Luigi Carbone, et.al., Nano Lett. 2007, 7, 10, 2942–2950 , https://doi.org/10.1021/nl0717661
47. Quantification of the Photon Absorption, Scattering, and On-Resonance Emission Properties of CdSe/CdS Core/Shell Quantum Dots: Effect of Shell Geometry and Volumes, Dongmao Zhang et. al Anal. Chem. 2020, 92, 7, 5346–5353, https://doi.org/10.1021/acs.analchem.0c00016
48. Vu Thi Kim Lien et al Optical properties of CdS and CdS/ZnS quantum dots synthesized by reverse micelle method. 2009 J. Phys.: Conf. Ser. 187 012028
49. Nor Aliya Hamizi, Mohd Rafie Johan Optical Properties of CdSe Quantum Dots via Non-TOP based Route Int. J. Electrochem. Sci., 7 (2012) 8458 – 8467

50. Wang, Y., Leck, K.S., Ta, V.D., Chen, R., Nalla, V., Gao, Y., He, T., Demir, H.V. and Sun, H. (2015), Blue Liquid Lasers from Solution of CdZnS/ZnS Ternary Alloy Quantum Dots with Quasi-Continuous Pumping. Adv. Mater., 27: 169-175. https://doi.org/10.1002/adma.201403237
51. Lin, W., Niu, Y., Meng, R. et al. Shell-thickness dependent optical properties of CdSe/CdS core/shell nanocrystals coated with thiol ligands. Nano Res. 9, 260–271 (2016). https://doi.org/10.1007/s12274-016-1014-0
52. Park, YS., Lim, J. & Klimov, V.I. Asymmetrically strained quantum dots with non-fluctuating single-dot emission spectra and subthermal room-temperature linewidths. Nature Mater 18, 249–255 (2019). https://doi.org/10.1038/s41563-018-0254-7
53. Fan, F. et al. Continuous-wave lasing in colloidal quantum dot solids enabled by facet-selective epitaxy. Nature 544, 75–79 (2017).
54. Lim, J., Park, Y.-S. & Klimov, V. I. Optical gain in colloidal quantum dots achieved with direct-current electrical pumping. Nat. Mater. 17, 42–49 (2018)
55. Colin D. Sonnichsen, Tobias Kipp, Xiao Tang, and Patanjali Kambhampati, Efficient Optical Gain in CdSe/CdS Dots-in-Rods ACS Photonics 2019 6 (2), 382-388, DOI: 10.1021/acsphotonics.8b01033
56. Riley, E.A.; Hess, C.M.; Reid, P.J. Photoluminescence Intermittency from Single Quantum Dots to Organic Molecules: Emerging Themes. Int. J. Mol. Sci. 2012, 13, 12487-12518. https://doi.org/10.3390/ijms131012487

## Ethics declarations

### Competing interests

The authors declare no competing financial interests.

## Supplementary Information